%% file: cosmo2k+idm2k-lr-net.tex
\documentclass[12pt]{article}
\usepackage{pictex,graphicx,color,rotating}
\newcommand{\newc}{\newcommand}
\newc\eg{{\it {e.g.}}}  \newc\etal{{\it {et al.}}} \newc\ie{{\it i.e.}}
\newc\etc{{\it etc}}
\newc\second{{\rm sec}} 

\newc\mone{M_1} 

\newc{\logsoms}{\log\left(s/m_{\rm eff}^2\right)}

\newc\alphas{\alpha_s}

\newc{\gstar}{g_\ast}           \newc{\gsstar}{g_{s\ast}}
\newc{\geff}{g_{\rm eff}}

\newcommand\mz{m_{Z}}
\newc\meff{m_{\rm eff}} 

\newc\sigmabar{{\overline{\sigma}}}

\newcommand\treh{T_{\rm R}}
   
\newcommand\Td{T_{\rm D}}
\newcommand\Teq{T_{\rm eq}}     
\newcommand\Tnr{T_{\rm NR}}
\newcommand\tf{T_{\rm f}}       
\newc{\xf}{x_f}

\newc{\nspin}{n_{\rm spin}}
\newc{\nflavor}{n_{\rm F}}

\newc\vrel{v_{\rm rel}}
\newcommand\fa{f_{a}}
\newcommand\mchi{m_{\chi}}

\newcommand\squark{\widetilde q}        \newcommand\msquark{m_{\squark}}

\newcommand\gluino{\widetilde g}
\newcommand\mgluino{m_{\widetilde g}}

\newcommand\axino{\tilde{a}}            \newcommand\maxino{m_{\axino}}

\newcommand\abunda{\Omega_{\axino}h^2}  \newcommand\omegaa{\Omega_{\axino}}

                \newcommand\omegaantp{\Omega^{\rm NTP}_{\axino}}
       
                \newcommand\omegaatp{\Omega^{\rm TP}_{\axino}}

\newc{\cachigamma}{C_{a\chi\gamma}}     \newc{\cachigammai}{C_{a\chi\gamma,i}}

\newc{\caww}{C_{aWW}}                   \newc{\cayy}{C_{aYY}}
\newc{\cagg}{C_{agg}}
\newc{\sthw}{\sin\theta_W}              \newc{\cthw}{\cos\theta_W}
\newc{\bino}{\widetilde B}              \newc{\wino}{\widetilde W_3}
\newc{\higgsinob}{{\widetilde H}^0_b}   \newc{\higgsinot}{{\widetilde H}^0_t}

\newc{\naxino}{n_{\tilde a}}
\newc{\ngamma}{n_\gamma}

\newc{\ychi}{Y_{\chi}}                  \newc{\yeqchi}{Y^{\rm EQ}_{\chi}}

\newc{\yaxino}{Y_{\axino}}
\newc{\yeqaxino}{Y^{\rm EQ}_{\axino}}
\newc{\ythaxino}{Y^{\rm TP}_{\axino}}
\newc{\ynthaxino}{Y^{\rm NTP}_{\axino}}
\newc{\yascat}{Y^{\rm scat}_{i,j}}      \newc{\yadec}{Y^{\rm dec}_{i}}
\newc{\abund}{\Omega h^2}
\newc{\abundchi}{\Omega_\chi h^2}
\newc{\rhocrit}{\rho_{crit}}
\newc{\rhochi}{\rho_{\chi}}
\newc{\rhoaxino}{\rho_{\axino}}
\newcommand\tev{\,\mbox{TeV}}
\newcommand\gev{\,\mbox{GeV}}
\newcommand\mev{\,\mbox{MeV}}
\newcommand\kev{\,\mbox{keV}}
\newcommand\ev{\,\mbox{eV}}

\newc{\mplanck}{M_{\rm P}}      \newc{\mpl}{M_{\rm Pl}}
\newc{\msusy}{M_{\rm SUSY}}      \newc{\ms}{M_{\rm S}}

\newc{\VEV}[1]{\langle #1 \rangle}

\newc{\ra}{\rightarrow}
\newc{\beq}{\begin{equation}}
\newc{\eeq}{\end{equation}}
\newc{\bea}{\begin{eqnarray}}
\newc{\eea}{\end{eqnarray}}

\newcommand\lsim{\mathrel{\rlap{\lower4pt\hbox{\hskip1pt$\sim$}}
    \raise1pt\hbox{$<$}}}
\newcommand\gsim{\mathrel{\rlap{\lower4pt\hbox{\hskip1pt$\sim$}}
    \raise1pt\hbox{$>$}}}

\begin{document}

\begin{titlepage}
\pagestyle{empty}
\baselineskip=21pt
\rightline{CERN--TH/2001-059}
\vskip 0.5in
\begin{center}
{\large {\bf Axino - New Candidate for Cold Dark Matter\footnote{Based
    on invited
plenary talks at the 4th International Workshop on Particle Physics and the
Early Universe (COSMO-2000), Cheju, Cheju Island, Korea, September
4--8, 2000
and the 3rd International Workshop on the Identification of Dark
Matter (IDM-2000), York, England, 18--22 September 2000.
}
}}
\end{center}
\begin{center}
\vskip 0.05in
{\bf Leszek Roszkowski}$^{1,2}$\\

\vskip 0.05in
{\it
$^1${Department of Physics, Lancaster University, Lancaster LA1
4YB, England}\\
$^2${TH Division, CERN, CH-1211 Geneva 23, Switzerland}\\
}
\vskip 0.5in
{\bf Abstract}
\end{center}
\baselineskip=18pt 
\noindent 
{\small Supersymmetric extensions of the Standard Model when combined
  with the Peccei-Quinn solution to the strong CP problem necessarily
  contain also the axino, the fermionic partner of the axion. In
  contrast to the neutralino and the gravitino, the axino mass is
  generically not of the order of the supersymmetry breaking scale and
  can be much smaller. The axino is therefore an intriguing candidate
  for a stable superpartner. The axinos are a natural candidate for
  cold dark matter in the Universe when they are generated
  non-thermally through out-of-equilibrium neutralino decays or via a
  competing thermal production mechanism through scatterings and
  decays of particles in the plasma. We identify axino masses in the
  range of tens of~MeV to several~GeV (depending on the scenario) as
  corresponding to cold axino relics if the reheating temperature
  $\treh$ is less than about $5\times10^4\gev$. At higher $\treh$ and
  lower mass, axinos could constitute warm dark matter. In the
  scenario with axinos as stable relics the gravitino problem finds a natural
  solution. The lightest superpartner of the Standard Model spectrum
  will remain stable in high-energy detectors but may be either
  neutral or charged. The usual constraint $\abundchi\lsim1$ on the
  relic abundance of the lightest neutralino does not  hold.}
\vfill
\vskip 0.15in
\leftline{CERN--TH/2001-059}
\leftline{February 2001}
\end{titlepage}
\baselineskip=15pt

\section{Introduction}
The nature of dark matter (DM) in the Universe remains unknown. Its
relic abundance is probably in the range $0.1\lsim\abundchi\lsim0.3$ or
so and is most likely ``cold''. It presumably consists for the most
part of some weakly interacting massive particles (WIMPs).

WIMPs do not necessarily have to interact via weak interactions {\em
  per se}. Very much weaker interaction strengths can also lead to
interesting cosmological abundances. One expects that they should
preferably be electrically and color neutral, and therefore be
non-baryonic, otherwise they would dissipate their kinetic energy.

Supersymmetry (SUSY) provides a natural context for WIMPs. In the presence
of $R$-parity, the (massive) lightest supersymmetric particle (LSP) is stable
and may substantially contribute to the relic mass density. It may therefore
constitute the DM in the Universe. 
Among SUSY WIMPs, the neutralino remains the most popular choice by
being perhaps the most ``genuine'' WIMP.  The gravitino is another
interesting candidate although it generically suffers from the (in)famous
``gravitino problem''.

Recently, it has been pointed out\cite{ckr,ckkr} that another
well-motivated SUSY particle can be a plausible cold DM (CDM)
candidate. The particle in question is the axino, a fermionic SUSY
partner of the axion. Axionic particles arise naturally in various
superstring models and, historically first, in the Peccei-Quinn (PQ)
solution to the strong CP problem. 

It is also well-justified to consider the axino as the LSP since its
mass is basically a free parameter which can only be determined in
specific models.  Since current LEP bounds on the neutralino are in
excess of 32~GeV, or so, it becomes increasingly plausible that there
may well be another SUSY particle which is lighter than the
neutralino, and therefore a candidate for the LSP and dark matter.
The axino is just such a candidate.

In this talk I will first describe the main properties of the new
candidate. Next I will discuss ways of producing axinos in the early
Universe and the expected abundance. Finally, I will make some remarks
about the ensuing implications for cosmology and phenomenology.

\section{Axino} 

Axino is a superpartner of the axion pseudoscalar.  It is thus a
neutral, $R$-parity negative, Majorana chiral fermion. There exist
several SUSY and supergravity implementations of the well-known
original axion models (KSVZ\cite{ksvz} and DFSZ\cite{dfsz}).
(Axion/axino-type supermultiplets also arise in superstring models.)
In studying cosmological properties of axinos, we will concentrate on
KSVZ and DFSZ-type models where the global $U(1)$ PQ symmetry is
spontaneously broken at the PQ scale $\fa$. A combination of
astrophysical (white dwarfs, \etc) and cosmological bounds gives
$10^9\gev\lsim\fa\sim10^{12}\gev$\cite{axionreviews:cite} although
the upper bound can be significantly relaxed if inflation followed the
decoupling of primordial axionic particles and the reheating
temperature $\treh\ll\fa$.

The two main parameters of interest to us are the axino mass and
coupling. The mass $\maxino$ strongly depends on an underlying model
and can span a wide range, from very small ($\sim\ev$) to large
($\sim\gev$) values. What is worth stressing is that, in contrast to the
neutralino and the gravitino, axino mass does not have to be of the
order of the SUSY breaking scale in the visible sector, $\msusy\sim
100\gev - 1\tev$\cite{tw,rtw,ckn}. The basic argument goes as
follows.  In the case of unbroken SUSY, all members of the axion
supermultiplet remain degenerate and equal to the tiny mass of the
axion given by the QCD anomaly.  Once SUSY is softly broken,
superpartners acquire mass terms. Since the axino is a chiral fermion,
one cannot write a dimension-four {\em soft} mass term for the axino.
(For the same reason there are no soft terms for, \eg, the MSSM
higgsinos.)  The lowest-order term one can write will be a
non-renormalizable term of dimension-five. The axino mass will then be
of order $\msusy^2/\fa\sim1\kev$\cite{tw,ckn}.

However, in specific SUSY models there are normally additional sources
of axino mass which can generate much larger contributions to
$\maxino$ at one-loop or even the tree-level.  One-loop terms will
always contribute but will typically be $\lsim\msusy$ (KSVZ) or even
$\ll\msusy$ (DFSZ).  Furthermore, in non-minimal models where the
axino mass eigenstate comes from more than one superfield, $\maxino$
arises even at the tree-level. In this case $\maxino$ can be of order
$\msusy$ but can also be much smaller. In a study of cosmological
properties of axinos, it therefore makes sense to treat their mass as
a basically free parameter.
 
Axino couplings to other particles are generically 
suppressed by $1/\fa$.
For our purpose the most important coupling will be
that of axino-gaugino-gauge boson interactions
which can written as a dimension-five term in the Lagrangian
\begin{equation}
{\cal L}_{\axino\lambda A} = i \frac{\alpha_Y \cayy}{16\pi\left(\fa/N\right)} 
{\bar{\axino}} \gamma_5 [\gamma^\mu,\gamma^\nu]\bino B_{\mu\nu}
+ i \frac{\alpha_s}{16\pi\left(\fa/N\right)}
{\bar{\axino}}\gamma_5[\gamma^\mu,\gamma^\nu]\gluino^b F^b_{\mu\nu},
\label{eq:axino-gaugino-gauge}
\end{equation}
where $\bino$ denotes the bino, the fermionic partner of the $U(1)_Y$
gauge boson $B$, which is one of the components of the neutralino, 
$\gluino$ stands for the gluino and $N=1(6)$ for the KSVZ
(DFSZ) model. One can show that the $SU(2)_L$
coupling can be rotated away so long as one discusses cosmological
properties of the axino at large temperatures. Depending on a model,
one can also think of terms involving dimension-four operators coming,
\eg, from the {\em effective} superpotential $\Phi\Psi\Psi$ where
$\Psi$ is one of MSSM matter (super)fields.  However, axino production
processes coming from such terms will be suppressed at high energies
relative to processes involving Eq.~(\ref{eq:axino-gaugino-gauge}) by
a factor $m_\Psi^2/s$ where $s$ is the square of the center of mass
energy.  We will comment on the role of dimension-four operators again
below but, for the most part, mostly concentrate on the processes
involving axino interactions with gauginos and gauge bosons,
Eq.~(\ref{eq:axino-gaugino-gauge}), which are both model-independent
and dominant.

\section{Axino Production} 
Particles like axinos and
gravitinos are somewhat special in the sense that their interactions
with other particles are very strongly suppressed compared to the
SM interaction strengths.  Therefore such particles remain
in thermal equilibrium only at very high temperatures. In the
particular case of axinos (and axions), their initial thermal
populations decouple at\cite{rtw} 
$\Td\sim 10^{10}\gev$.
At such high temperatures, the axino co-moving number density is the
same as the one of photons.  In other
words, such primordial axinos freeze out as {\it relativistic}
particles. Rajagopal, Turner and Wilczek (RTW)\cite{rtw} pointed out
that, in the absence of a subsequent period of inflation, the
requirement that the axino energy density is not too large
($\omegaa\lsim1$) leads to $\maxino < 2\kev$ 
and the corresponding axinos would be light and would provide
warm or even 
hot dark matter\cite{rtw}. We will not consider this case in the
following primarily because we are interested in cold DM axinos.  We
will therefore assume that the initial population of axinos (and other
relics, like
gravitinos) present in the early Universe was subsequently diluted
away by an intervening inflationary stage and that the reheating
temperature after inflation was smaller than $\Td$. It also had to be
less than $\fa$, otherwise the PQ would have been restored thus
leading to the well-known domain wall problem associated with global
symmetries. 

In order to generate large enough abundance of axinos, one needs to
repopulate the Universe with them.  There are two generic ways of
doing achieving this.  First, they can be generated through thermal
production, namely via two-body scattering and decay processes of
ordinary particles and sparticles still in thermal bath.  (Despite the
name of the process, the resulting axinos will typically be already
out-of-equilibrium because of their exceedingly tiny couplings to
ordinary matter.)  Second, axinos may also be produced in decay
processes of particles which themselves are out-of-equilibrium. Such
particles could for example be ordinary superpartners, the gravitino
or the inflaton field.  (By ``ordinary'' we mean a particle or its
superpartner carrying only Standard Model quantum numbers.) Below we
will concentrate on the first possibility in the case the decaying
neutralino.

\subsection{Thermal Production} 

After inflation the Universe can be re-populated with
axinos (and gravitinos) through scattering and decay processes
involving superpartners in thermal bath.  As long as the axino
co-moving number density $\naxino$ is much smaller than $\ngamma$, the
number density of photons in thermal equilibrium, its time evolution
will be adequately described by the Boltzmann equation
\begin{equation}
\label{eq:Boltzmann}
\frac{d\,\naxino}{d\,t} + 3H \naxino =
\sum_{i,j} \langle\sigma(i+j\ra \axino+\cdots)\vrel\rangle n_i n_j +
\sum_i \langle\Gamma(i\ra \axino+\cdots)\rangle n_i.
\end{equation}
Here $H$ is the Hubble parameter,
$H(T)=\sqrt{\frac{\pi^2\gstar}{90\mplanck^2}}\;T^2$, 
where $g_*$ is the effective relativistic degrees of freedom,
$\sigma(i+j\ra \axino+\cdots)$ is the scattering cross section for
particles $i,j$ into final states involving axinos,
$\vrel$ is their relative velocity,
$n_i$ is the $i$th particle number density in thermal bath,
$\Gamma(i\ra \axino+\cdots)$ is the decay width of the $i$th particle and
$\langle\cdots\rangle$ stands for thermal averaging. (Averaging over initial
spins and summing over final spins is understood.)
Note that on the {\it r.h.s.} we have neglected inverse processes
since they are suppressed by $\naxino$.

The main axino production channels are the scatterings of
(s)particles described by a dimension-five axino-gaugino-gauge boson
term in the Lagrangian~(\ref{eq:axino-gaugino-gauge}).  Because of the
relative strength of $\alphas$, the most important contributions will
come from 2-body strongly interacting processes into final states
involving axinos, $i+j\ra \axino+\cdots$. (Scattering processes involving
electroweak interactions are suppressed by both the strength of the
coupling and a smaller number of production channels\cite{ckkr}.)
The cross section can be written as
\begin{equation}
\label{eq:cross-section}
\sigma_n(s) = \frac{\alphas^3}{4\pi^2\left(\fa/N\right)^2}{\sigmabar}_n(s)
\end{equation}
where $\sqrt{s}$ is the center of mass energy and 
$n=A,\ldots,K$ refers to different channels which are 
listed in Table~I in Ref.~\cite{ckkr}. 
The diagrams listed in the Table are analogous to those involving
gravitino production and  we use the same classification.  
This analogy should not be surprising since both particles are neutral
Majorana superpartners. 

In addition to scattering
processes, axinos can also be produced through decays of heavier
superpartners in thermal plasma. At temperatures $T\gsim\mgluino$
these are dominated by the decays of gluinos into LSP axinos and
gluons.  The relevant decay width is given by
\begin{equation}
\label{eq:Gamma-gluino}
\Gamma(\gluino^a\ra \axino + g^b) = \delta^{ab}
\frac{\alpha_s^2}{128\pi^3}\frac{\mgluino^3}{\left(\fa/N\right)^2}\left(1-
\frac{\maxino^2}{\mgluino^2}\right)^3
\end{equation}
and one should sum over the color index $a,b=1,\cdots,8$.
At lower temperatures $\mchi\lsim \treh\lsim\mgluino$, neutralino
decays to axinos also contribute while at higher temperatures they are
sub-dominant. They only become important when the axino yield 
\begin{equation}
\ythaxino = \frac{\naxino^{\rm TP}}{s},
\label{ythaxino:eq}
\end{equation}
where 
$s= (2\pi^2/45)\gsstar T^3$ is the entropy density, 
and normally $\gsstar=\gstar$ in the early Universe, becomes too small
to be cosmologically interesting.

The results are presented in Fig.~1 
for representative values of $\fa=10^{11}\gev$ and $\msquark=\mgluino=1\tev$.
The respective contributions due to scattering as well as gluino and
neutralino decays are marked by dashed, dash-dotted and dotted lines.

\begin{figure}[t]
\label{fig:YTP-TR}
\include{ckkr-fig1}
\caption{$\ythaxino$ as a function of $\treh$ for representative
values of $\fa=10^{11}$GeV and $\msquark = \mgluino = 1\tev$.}
\end{figure}

It is clear that at high enough $\treh$, much above $\msquark$ and
$\mgluino$, scattering processes involving such particles dominate the
axino production.  For $\treh\gg m_{\tilde q},m_{\tilde g}$, 
$Y^{\rm scat}$ grows linearly as $\treh$ becomes larger.  
In contrast, the decay contribution above the gluino
mass threshold, $Y^{\rm dec}\simeq 5\times10^{-4}\,(M_P\Gamma_{\tilde
g}/m_{\tilde g}^2)$, remains independent of $\treh$.
At $\treh$ roughly below the mass of the squarks and gluinos, their
thermal population starts to become strongly suppressed by the
Boltzmann factor $e^{-m/T}$, hence a distinct knee in the scattering
contribution in Fig.~1. 
It is in this region that gluino decays
(dash-dotted line) given by Eq.~(\ref{eq:Gamma-gluino}) become
dominant before they also become suppressed by the Boltzmann factor
due to the gluino mass.
For $\mchi\lsim \treh\lsim m_{\tilde q},m_{\tilde g}$,
the axino yield is well approximated by
$Y^{\rm TP}\approx Y^{\rm dec}\simeq
5\times10^{-4}(M_P\Gamma_{\tilde g}/\treh^2)\, e^{-m_{\tilde g}/\treh}$,
and depends sensitively on the reheating temperature.

At still lower temperatures the population of strongly interacting
sparticles becomes so tiny that at $\treh\sim\mchi$ neutralino decays
start playing some role before they
too become Boltzmann factor suppressed. We indicate this by plotting
in Fig.~1 
the contribution of the lightest neutralino (dotted line).  It is
clear that the values of $\ythaxino$ in this region are so small that,
as we will see later, they will not play any role in further
discussion. We therefore do not present the effect of the decay of the
heavier neutralinos.  Furthermore, model-dependent dimension-four
operators will change axino production cross section at lower
$\treh\sim\msusy$ but will be suppressed at high temperatures. We have
not studied this point yet.

We emphasize that axinos produced in this way are already out of
equilibrium. Their number density is very much smaller than $\ngamma$
(except  $\treh\sim10^9\gev$ and above) 
and cross sections for axino re-annihilation into other particles are
greatly suppressed. This is why in Eq.~(\ref{eq:Boltzmann}) we have neglected
such processes. Nevertheless, even though axinos never reach equilibrium, 
their number density may be large enough to give $\omegaa\sim 1$ for
large enough axino masses ($\kev$ to $\gev$ range) as we will see
later.

\subsection{Non-Thermal Production} 
The mechanism for non-thermal production (NTP) that we will consider
works as follows.  Consider some lightest ordinary superpartner
(LOSP). Because axino LSP couplings to everything else are suppressed
by $1/\fa$, as the Universe cools down, all heavier SUSY partners will
first cascade-decay to the LOSP. The LOSPs then freeze out of thermal
equilibrium and subsequently decay into axinos.

A natural (although not unique), candidate for the LOSP is the
lightest neutralino. 
For example, 
in models employing full unification of
superpartner masses (like the CMSSM/mSUGRA), a mechanism of
radiative electroweak symmetry breaking typically
implies $\mu^2\gg\mone^2$, where $\mone$ is the 
bino mass parameter. As a result, the bino-like neutralino often
emerges as the lightest ordinary superpartner\cite{na93,rr93,kkrw}.

In the following we will assume that LOSP is the neutralino. 
It can decay to the axino and photon $\chi\ra\axino\gamma$ with 
the rate\cite{ckr,ckkr}
\beq
\Gamma(\chi\ra\axino\gamma) = 
{\alpha^2_{em} \cachigamma^2 \over 128 \pi^3} 
{{\mchi}^3\over {\left(\fa/N\right)^2}} \left(1-{\maxino^2 \over
    \mchi^2}\right)^3.
\label{gammachi:eq}
\eeq
Here $\alpha_{em}$ is the electromagnetic coupling strength,
$\cachigamma=(\cayy/\cos\theta_W) Z_{11}$, with $Z_{11}$ standing for
the bino part of the lightest neutralino. (We use
the basis $\chi_i= Z_{i1}\bino + Z_{i2}\wino +
Z_{i3}\higgsinob + Z_{i4}\higgsinot$ ($i=1,2,3,4$) of the respective fermionic
partners (denoted by a tilde) of the electrically neutral gauge bosons 
$B$ and $W_3$, and the MSSM Higgs bosons $H_b$ and $H_t$.)

The corresponding lifetime can be written as
\beq
\tau(\chi\ra\axino\gamma)=  {\frac{0.33\, {\rm sec}}{\cayy^2 Z_{11}^2}}
\left({\frac{\alpha^2_{em}}{1/128}}\right)^{-2}
\left(\frac{f_a/N}{10^{11}\gev}\right)^2
\left(\frac{100\gev}{\mchi}\right)^3 
\left(1-{\maxino^2 \over \mchi^2}\right)^{-3}.
\label{chilife:eq}
\eeq

For large enough neutralino mass, an additional 
decay channel into axino and Z boson opens up but 
is always subdominant relative to
$\chi\ra\axino\gamma$ because of both the phase-space suppression and
the additional factor of $\tan^2\theta_W$. As a result, even at
$\mchi\gg \mz,\maxino$, $\tau(\chi\ra\axino Z)\simeq 
3.35\, \tau(\chi\ra\axino\gamma)$. 
It is also clear that the neutralino lifetime rapidly decreases with its
mass ($\sim1/\mchi^3$).  On the other hand, if the neutralino is not
mostly a bino, its decay will be suppressed by the $Z_{11}$ - factor
in $\cachigamma$. 

Other decay channels are the decay into axino and Standard Model fermion 
pairs through virtual photon or $Z$ but they are negligible compared
with the previous ones. We will discuss them later since, for a low
neutralino mass, \ie, long lifetime, they can, even if subdominant,
produce dangerous hadronic showers during and after nucleosynthesis.

Additionally, in the DFSZ type of models, there exists an additional
Higgs-higgsino-axino couplings, which may open up other
channels\cite{ckkr}. These are model-dependent and I will not discuss
them here.

\subsection{Constraints} 

Several nontrivial conditions have to be satisfied in order for
axinos to be a viable CDM candidate. First, we will expect their relic
abundance to be large enough, $\abunda\simeq0.2$. This obvious
condition will have strong impact on other bounds. Next, the
axinos generated through both TP and NTP will in most cases be
initially relativistic. We will therefore require that they become
non-relativistic, or cold, much before the era of matter
dominance.  Furthermore, since NTP axinos will be produced near the
time of BBN, we will require that they do not contribute too much
relativistic energy density to radiation during BBN. Finally,
axino production associated decay products will often result in
electromagnetic and hadronic showers which, if too large, would cause
too much destruction of light elements. In deriving all of these
conditions, except for the first one, the lifetime of the parent LOSP
will be of crucial importance.

A detailed discussion of the bounds would take too much time,
and space, that is available. I will therefore merely summarize the
relevant results. 
First, the condition that the axinos give a dominant
contribution to the matter density at the present time can be expressed as 
$
\maxino \yaxino \simeq 0.72 \ev \left(\abunda/ 0.2\right)
$
which  applies to both TP and NTP relics. It is worth mentioning here
that, for the initial population of axinos, the
yield at decoupling is approximately
$\yaxino\simeq\yeqaxino\simeq2\times10^{-3}$ which gives 
$\maxino \simeq 0.36 \kev \left(\abunda/ 0.2\right)$. 
This is an updated value for the RTW
bound.
Next, we want to determine the temperature of the Universe at which
the axinos will become non-relativistic. 
In nearly all cases axinos  are initially relativistic
and, due to expansion, become non-relativistic at some later epoch
which depends on their mass and  production mechanism. 
In the case of TP, axinos are not in thermal equilibrium but, since they are
produced in kinetic equilibrium with the thermal bath, their momenta
will have a thermal spectrum.  They will
become non-relativistic when the thermal bath temperature reaches the
axino mass, $\Tnr\simeq \maxino$.

NTP axinos generated through out-of-equilibrium neutralino decays
will be produced basically monochromatically, all
with the same energy roughly given by $\mchi/2$, unless they are
nearly mass-degenerate with the neutralinos. This is so because the
neutralinos, when they decay, are themselves already non-relativistic.
Thus, due to momentum red-shift, 
axinos will become non-relativistic only at a later
time, when $p(\Tnr) \simeq \maxino$. The
temperature $\Tnr$ can be expressed as\cite{ckkr} 
\begin{equation}
\Tnr  = 4.2\times 10^{-5}\maxino \cayy Z_{11}
\left( {\mchi/ 100 \gev} \right)^{1/2}
\left({10^{11} \gev/f_a/N}\right).
\end{equation}

This epoch has to be compared to the matter-radiation equality
epoch given by $\Teq = 1.1 \ev \left(\abunda/ 0.2\right)$
which holds for both thermal and non-thermal production.
In the TP case one can
easily see that $\Tnr>\Teq$ is satisfied for any interesting range of
$\maxino$.  In the case of NTP the condition $\Tnr\gg\Teq$ is satisfied for 
\bea
\maxino &\gg & 27 \kev {1\over \cayy Z_{11}} 
\left( {100 \gev \over \mchi} \right)^{1/2} 
\left({ f_a/N\over 10^{11} \gev}\right)
\left(\abunda \over 0.2\right).
\label{eq:relaxinobbnone}
\eea
If axinos were lighter than the bound~(\ref{eq:relaxinobbnone}), 
then the point of radiation-matter equality would be shifted
to a later time around $\Tnr $. Note that in this case axino
would not constitute cold, but warm or hot dark matter.
In the NTP case discussed here
other constraints would however require the axino 
mass to be larger than the above bound, so that we can discard this 
possibility.

BBN predictions provide further important
constraints on axinos as relics. 
In the case of non-thermal production most axinos will be produced
only shortly before nucleosynthesis and, being still
relativistic, may dump too much to the energy density during
the formation of light elements. 
In order not to affect the Universe's expansion during BBN, axino
contribution to the energy density should satisfy
$
{\rhoaxino / {\rho_{\nu}}} \leq {\delta N_{\nu}},
$
where $\rho_{\nu}$ is the energy density of one
neutrino species. Agreement with observations of light elements 
requires
${\delta N_{\nu}} = 0.2 - 1$. This leads to\cite{ckkr}
\bea
\maxino &\gsim & 181  \kev {1\over {\delta N_{\nu}}} {1\over \cayy Z_{11}} 
\left( {100 \gev \over \mchi} 
\right)^{1/2} \left({ f_a/N\over 10^{11} \gev}\right)
\left(\abunda \over 0.2\right).
\eea

Finally, photons and quark-pairs produced in NTP decays of
neutralinos, if
produced during or after BBN, may lead to a significant depletion of
primordial elements. One often applies a crude constraint
that the lifetime should be less than about 1~second which in our case
would provide a lower bound on $\mchi$. 
First, photons produced in reaction $\chi\ra\axino\gamma$ carry a
large amount of energy, roughly $\mchi/2$. If the decay takes place
before BBN, the photon will rapidly thermalize via multiple
scatterings from background electrons and positrons. The process will be
particularly efficient at plasma temperatures above $1\mev$ which is
the threshold for background $e\bar e$ pair annihilation, and
which, incidentally, coincides with time of about 1~second. But a
closer examination\cite{bbnbound} 
shows that also scattering with the high energy
tail of the CMBR thermalize photons very efficiently and so the decay
lifetime into photons can be as large as $10^4$~sec. By comparing this
with Eq.~(\ref{chilife:eq}) we find that, in the gaugino regime, this
can be easily satisfied for $\mchi<\mz$. It is only in a
nearly pure higgsino case and mass of tens of $\gev$ that the bound
would become constraining. We are not interested in such light
higgsinos for other reasons, as will be explained later.

A much more stringent constraint comes from considering hadronic
showers from $q\bar q$-pairs. These will be produced through a virtual
photon and $Z$ exchange, and, above the kinematic threshold for
$\chi\ra\axino Z$, also through the exchange of a real $Z$-boson.
Here the discussion is somewhat more involved and the resulting
constraint strongly depends on $\mchi$. One can show\cite{ckkr} that
at the end one gets
roughly $\maxino\gsim 360\mev$ for $\mchi\lsim60\gev$ which gives the
strongest bound so far. However, the bound on $\maxino$ decreases
nearly linearly with $\mchi$ and disappears completely for
$\mchi\gsim150\gev$. 

In summary, a lower bound $\maxino\gsim{\cal O}(300\kev)$ arises from
either requiring the axinos to be cold at the time of matter dominance
or that they do not contribute too much to the relativistic energy
density during BBN. The constraint from hadronic destruction of light
elements can be as strong as $\maxino\gsim 360 \mev$ (in the
relatively light bino case) but it is highly model-dependent and
disappears for larger $\mchi$.

\begin{figure}[t]
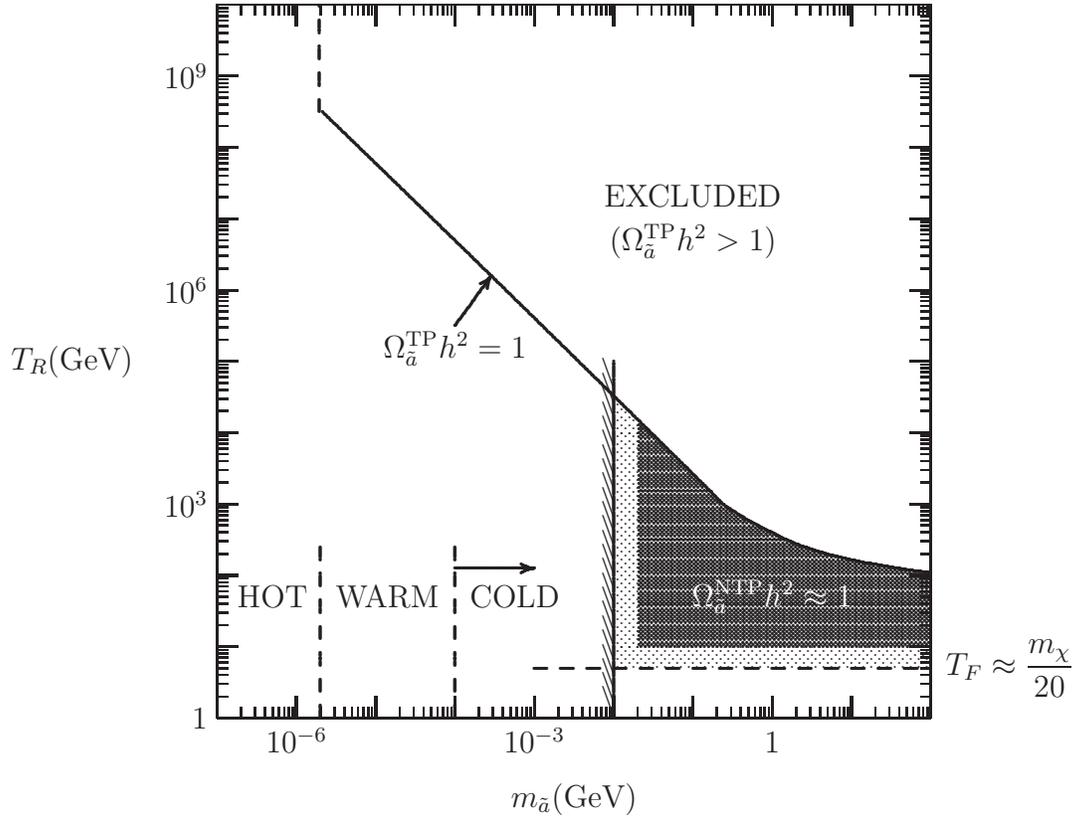

\label{fig:TR-maxino}
\include{ckkr-fig3}
\caption{
The thick solid line gives the upper bound from thermal production
on the reheating temperature as a function of the axino mass.  The
dark region is the region where non-thermal production can give
cosmologically interesting results 
($\Omega^{\rm NTP}_{\axino} h^2\simeq1$) as explained in the text.  
We assume a bino-like neutralino with $\mchi=100\gev$
and $\fa=10^{11}\gev$. The region of $\treh\gsim\tf$ is somewhat
uncertain and has been denoted with light-grey color. A sizeable
abundance of neutralinos (and therefore axinos) is expected also for
$\treh\lsim\tf$ but has not been
calculated. The vertical light-grey band indicates that a low range of
$\maxino$ corresponds to allowing SM superpartner masses in the
multi-TeV range, as discussed in the text.
Division of hot, warm and cold dark matter by the axino mass shown
in lower left part is for axinos from non-thermal production.
}
\end{figure}

\subsection{Relic Abundance from Thermal and Non-Thermal Production} 

In the TP case the axino yield is
primarily determined by the reheating temperature.
For large enough $\treh$ ($\treh\gg\mgluino,\msquark$), it is
proportional to $\treh/\fa^2$.
In contrast, the NTP axino yield is for the most part
independent of $\treh$ (so long as
$\treh\gg\tf$, the neutralino freezeout temperature). 
In the NTP case, the yield of axinos is
just the same as that of the decaying neutralinos. This leads
to\cite{ckr} 
$
\abunda = {\maxino/\mchi}\, \abundchi$, where
$\abundchi$ stands for the abundance that the
neutralinos would have had today had they not decayed into axinos.

In order to be able to compare both production
mechanisms, we will therefore fix the neutralino mass at some typical
value. Furthermore we will map out a cosmologically interesting range
of axino masses for which $\omegaantp\sim1$.
Our results are presented in Fig.~2 in the case of a nearly pure bino.
We also fix $\mchi=100\gev$ and $\fa=10^{11}\gev$. The dark region is
derived in the following way. It is well known that $\abundchi$ 
can take a wide range of values
spanning several orders of magnitude. In the framework of the MSSM,
which we have adopted, global scans give $\abundchi\lsim10^4$ in the
bino region at $\mchi\lsim100\gev$.
(This limit decreases roughly linearly (on a log-log scale) down to
$\sim10^3$ at $\mchi\simeq400\gev$.)
For $\mchi=100\gev$
we find that the expectation $\Omega^{\rm NTP}_{\axino} h^2\simeq1$ gives 
\begin{equation}
\label{ntpaxinorange:eq}
10\mev\lsim\maxino\lsim\mchi.
\end{equation}
We note, however, that the upper bound $\abundchi\lsim10^4$ comes from
allowing very large $\msusy$ (\ie, sfermion and heavy Higgs masses) in
the range of tens of $\tev$. Restricting all SUSY mass parameter below
about $1\tev$ reduces $\abundchi$ below $10^2$ and, accordingly,
increases the lower bound $\maxino\gsim1\gev$. Still, for the sake of
generality, we have the much more generous
bound~(\ref{ntpaxinorange:eq}) in Fig.~2 but marked a low range of
$\maxino$ with a light grey band to indicate the above point. 

Likewise, for reheating temperatures just above $\tf$ standard
estimates of $\abundchi$ become questionable. We have therefore
indicated this range of $\treh$ with again a light grey color. 
It has also been recently pointed out\cite{gkr00} that even in the
case of very low reheating temperatures $\treh$ below the LOSP
freezeout temperature, a significant population of them will be
generated during the reheating phase. Such LOSPs would then also decay
into axinos as above. We have not considered such cases in our
analysis and accordingly left the region $\treh<\tf$ blank even though
in principle we would expect some sizeable range of $\abunda$ there.

We can see that for large $\treh$, the TP mechanism is more important
than the NTP one, as expected.  Note also that in the TP case the
cosmologically favored region ($0.2\lsim\abunda\lsim0.4$) would form a
very narrow strip (not indicated in Fig.~2) just below the
$\omegaatp=1$-boundary.  In contrast, the NTP mechanism can give the
cosmologically interesting range of the axino relic abundance for a
relatively wide range of $\maxino$ so long as $\treh\lsim 5\times
10^4\gev$. Perhaps in this sense, the NTP mechanism can be considered
as somewhat more robust.

\subsection{Conclusions} 

The intriguing possibility that the axino is the LSP and dark matter
possesses a number of very distinct features which makes this case very
different from those of both the neutralino and the gravitino. In
particular, the axino can be a cold DM WIMP for a rather wide range of
masses in the $\mev$ to $\gev$ range and for relatively low reheating
temperatures $\treh\lsim5\times10^4\gev$. As $\treh$ increases,
thermal production of axinos starts dominating over non-thermal
production and the axino typically becomes a warm DM relic with a mass
broadly in a $\kev$ range.  In contrast, the neutralino is typically a
cold DM WIMP.

Low reheating temperatures would favor baryogenesis at the electroweak
scale. It would also alleviate the nagging ``gravitino
problem''. If additionally it is the axino that is the LSP and the
gravitino is the NLSP, the gravitino problem is resolved altogether
for both low and high $\treh$. 

Phenomenologically, one faces a well-justified possibility that the
bound $\abundchi<1$, which is often imposed in constraining a SUSY
parameter space, may be readily avoided. In fact, the range
$\abundchi\gg 1$ (and with it typically large masses of superpartners)
would now be favored if axino is to be a dominant component of DM in
the Universe.  Furthermore, the lightest ordinary superpartner could
either be neutral or charged but would appear stable in collider
searches.

The axino, with its exceedingly tiny coupling to other matter, will
be a real challenge to experimentalist. It is much more plausible that
a supersymmetric particle and the axion will be found first. Unless
the neutralino (or some other WIMP) is detected in DM searches, the
axino will remain an attractive and robust candidate for solving the
outstanding puzzle of the nature of dark matter in the Universe.


\end{document}

%% file: ckkr-fig1.tex
%

$$\beginpicture
\setcoordinatesystem units <27pt,18pt> point at 0 0
\setplotarea  x from 0.0 to 10.0, y from -15.0 to 0.0
\inboundscheckon
\linethickness 0.5pt

\axis bottom label {{$T_R$(GeV)}}
	ticks in
	width <0.8pt> length <8.0pt>
	withvalues {$1$} {$10^3$} {$10^6$} {$10^9$} /
	at 0 3 6 9 /
	width <0.8pt> length <8.0pt>
	at 1 2 4 5 7 8 /
       	width <0.5pt> length <4.0pt> at
       	0.301 0.477 0.602 0.699 0.778 0.845 0.903 0.954
       	1.301 1.477 1.602 1.699 1.778 1.845 1.903 1.954
       	2.301 2.477 2.602 2.699 2.778 2.845 2.903 2.954
       	3.301 3.477 3.602 3.699 3.778 3.845 3.903 3.954
       	4.301 4.477 4.602 4.699 4.778 4.845 4.903 4.954
       	5.301 5.477 5.602 5.699 5.778 5.845 5.903 5.954
       	6.301 6.477 6.602 6.699 6.778 6.845 6.903 6.954
       	7.301 7.477 7.602 7.699 7.778 7.845 7.903 7.954
       	8.301 8.477 8.602 8.699 8.778 8.845 8.903 8.954
       	9.301 9.477 9.602 9.699 9.778 9.845 9.903 9.954
       	/
/

\axis top label {}
	ticks in
	width <0.8pt> length <8.0pt>
	at 0 1 2 3 4 5 6 7 8 9 /
       	width <0.5pt> length <4.0pt> at
       	0.301 0.477 0.602 0.699 0.778 0.845 0.903 0.954
       	1.301 1.477 1.602 1.699 1.778 1.845 1.903 1.954
       	2.301 2.477 2.602 2.699 2.778 2.845 2.903 2.954
       	3.301 3.477 3.602 3.699 3.778 3.845 3.903 3.954
       	4.301 4.477 4.602 4.699 4.778 4.845 4.903 4.954
       	5.301 5.477 5.602 5.699 5.778 5.845 5.903 5.954
       	6.301 6.477 6.602 6.699 6.778 6.845 6.903 6.954
       	7.301 7.477 7.602 7.699 7.778 7.845 7.903 7.954
       	8.301 8.477 8.602 8.699 8.778 8.845 8.903 8.954
       	9.301 9.477 9.602 9.699 9.778 9.845 9.903 9.954
       	/
/

\axis left label {{$Y^{\rm TP}_{\tilde a}$}}
	ticks in
	width <0.8pt> length <8.0pt>
	withvalues {$1$} {$10^{-3}$} {$10^{-6}$} {$10^{-9}$} {$10^{-12}$} {$10^{-15}$} /
	at 0 -3 -6 -9 -12 -15 /
	width <0.8pt> length <8.0pt>
	at -1 -2 -4 -5 -7 -8 -10 -11 -13 -14 /
       	width <0.5pt> length <4.0pt> at
	-0.699 -0.523 -0.398 -0.301 -0.222 -0.155 -0.097 -0.046
	-1.699 -1.523 -1.398 -1.301 -1.222 -1.155 -1.097 -1.046
	-2.699 -2.523 -2.398 -2.301 -2.222 -2.155 -2.097 -2.046
	-3.699 -3.523 -3.398 -3.301 -3.222 -3.155 -3.097 -3.046
	-4.699 -4.523 -4.398 -4.301 -4.222 -4.155 -4.097 -4.046
	-5.699 -5.523 -5.398 -5.301 -5.222 -5.155 -5.097 -5.046
	-6.699 -6.523 -6.398 -6.301 -6.222 -6.155 -6.097 -6.046
	-7.699 -7.523 -7.398 -7.301 -7.222 -7.155 -7.097 -7.046
	-8.699 -8.523 -8.398 -8.301 -8.222 -8.155 -8.097 -8.046
	-9.699 -9.523 -9.398 -9.301 -9.222 -9.155 -9.097 -9.046
	-10.699 -10.523 -10.398 -10.301 -10.222 -10.155 -10.097 -10.046
	-11.699 -11.523 -11.398 -11.301 -11.222 -11.155 -11.097 -11.046
	-12.699 -12.523 -12.398 -12.301 -12.222 -12.155 -12.097 -12.046
	-13.699 -13.523 -13.398 -13.301 -13.222 -13.155 -13.097 -13.046
	-14.699 -14.523 -14.398 -14.301 -14.222 -14.155 -14.097 -14.046
       	/
/

\axis right label {}
	ticks in
	width <0.8pt> length <8.0pt>
	at 0 -1 -2 -3 -4 -5 -6 -7 -8 -9 -10 -11 -12 -13 -14 -15 /
       	width <0.5pt> length <4.0pt> at
	-0.699 -0.523 -0.398 -0.301 -0.222 -0.155 -0.097 -0.046
	-1.699 -1.523 -1.398 -1.301 -1.222 -1.155 -1.097 -1.046
	-2.699 -2.523 -2.398 -2.301 -2.222 -2.155 -2.097 -2.046
	-3.699 -3.523 -3.398 -3.301 -3.222 -3.155 -3.097 -3.046
	-4.699 -4.523 -4.398 -4.301 -4.222 -4.155 -4.097 -4.046
	-5.699 -5.523 -5.398 -5.301 -5.222 -5.155 -5.097 -5.046
	-6.699 -6.523 -6.398 -6.301 -6.222 -6.155 -6.097 -6.046
	-7.699 -7.523 -7.398 -7.301 -7.222 -7.155 -7.097 -7.046
	-8.699 -8.523 -8.398 -8.301 -8.222 -8.155 -8.097 -8.046
	-9.699 -9.523 -9.398 -9.301 -9.222 -9.155 -9.097 -9.046
	-10.699 -10.523 -10.398 -10.301 -10.222 -10.155 -10.097 -10.046
	-11.699 -11.523 -11.398 -11.301 -11.222 -11.155 -11.097 -11.046
	-12.699 -12.523 -12.398 -12.301 -12.222 -12.155 -12.097 -12.046
	-13.699 -13.523 -13.398 -13.301 -13.222 -13.155 -13.097 -13.046
	-14.699 -14.523 -14.398 -14.301 -14.222 -14.155 -14.097 -14.046
       	/
/

\setplotsymbol (.)

\setquadratic

%
%

\setsolid
\plot
0.5  -23.33458927564866
0.6  -20.749762339472287
0.7  -18.743247126148205
0.8  -17.19489313918617
0.9  -16.008966268937304
1.0  -15.109068781192619
1.1  -14.434110478058159
1.2  -13.935132988739149
1.3  -13.572815906128026
1.4  -13.315499567476689
1.5  -13.137607818479344
1.6  -13.018447399858548
1.7  -12.941393450089734
1.8  -12.593386502284663
1.9000000000000008  -11.652400901111502
2.000000000000001  -10.744575327186427
2.100000000000001  -10.05819685711149
2.200000000000001  -9.542625062631316
2.300000000000001  -9.156216525150487
2.4000000000000012  -8.863949184432531
2.5000000000000013  -8.636662181279513
2.6000000000000014  -8.451341734156493
2.7	-8.27
2.8	-8.09
2.9	-7.95
3.  -7.834430281105249
3.1  -7.743615715613835
3.2  -7.652773980744965
3.3000000000000003  -7.561902813174612
3.4000000000000004  -7.4710001864982285
3.5000000000000004  -7.380064288881011
3.6000000000000005  -7.289093503006911
3.7000000000000006  -7.198086388062038
3.8000000000000007  -7.107041663521974
3.900000000000001  -7.015958194541682
4.000000000000001  -6.924834978771702
4.1000000000000005  -6.833671134445942
4.2  -6.742465889605033
4.3  -6.6512185723354
4.3999999999999995  -6.559928601918263
4.499999999999999  -6.468595480794991
4.599999999999999  -6.377218787265938
4.699999999999998  -6.285798168849197
4.799999999999998  -6.194333336233837
4.899999999999998  -6.102824057769483
4.999999999999997  -6.011270154440216
5.099999999999997  -5.9196714952765115
5.199999999999997  -5.828027993163677
5.299999999999996  -5.736339601009678
5.399999999999996  -5.644606308239029
5.499999999999996  -5.552828137582821
5.599999999999995  -5.461005142137993
5.699999999999995  -5.369137402671623
5.7999999999999945  -5.277225025148399
5.899999999999994  -5.185268138461592
5.999999999999994  -5.093266892349669
6.099999999999993  -5.001221455482535
6.199999999999993  -4.909132013702753
6.299999999999993  -4.816998768408545
6.399999999999992  -4.724821935066616
6.499999999999992  -4.632601741843886
6.599999999999992  -4.54033842834824
6.699999999999991  -4.448032244469302
6.799999999999991  -4.35568344931101
6.899999999999991  -4.263292310208558
6.99999999999999  -4.170859101822839
7.09999999999999  -4.078384105306211
7.1999999999999895  -3.9858676075338666
7.299999999999989  -3.8933099003956153
7.399999999999989  -3.8007112801433207
7.4999999999999885  -3.708072046789643
7.599999999999988  -3.6153925035540793
7.699999999999988  -3.522672956352632
7.799999999999987  -3.4299137133277737
7.899999999999987  -3.3371150844155926
7.999999999999987  -3.244277380947298
8.099999999999987  -3.1514009152824674
8.199999999999987  -3.0584860004716443
8.299999999999986  -2.9655329499460734
8.399999999999986  -2.872542077232554
8.499999999999986  -2.779513695691488
/

\setdashes <3pt>
\plot
1.7000000000000008  -17.767610676653104
1.8000000000000008  -15.367610676653104
1.9000000000000008  -12.967610676653104
2.000000000000001  -11.927619096433935
2.100000000000001  -11.104638370560801
2.200000000000001  -10.449278078633878
2.300000000000001  -9.922584410821651
2.4000000000000012  -9.492265944383666
2.5000000000000013  -9.13486323870236
2.6000000000000014  -8.834368746345922
2.7	-8.53
2.8	-8.23
2.9	-7.95
3.  -7.834430281105249
3.1  -7.743615715613835
3.2  -7.652773980744965
3.3000000000000003  -7.561902813174612
3.4000000000000004  -7.4710001864982285
3.5000000000000004  -7.380064288881011
3.6000000000000005  -7.289093503006911
3.7000000000000006  -7.198086388062038
3.8000000000000007  -7.107041663521974
3.900000000000001  -7.015958194541682
4.000000000000001  -6.924834978771702
4.1000000000000005  -6.833671134445942
4.2  -6.742465889605033
4.3  -6.6512185723354
4.3999999999999995  -6.559928601918263
4.499999999999999  -6.468595480794991
4.599999999999999  -6.377218787265938
4.699999999999998  -6.285798168849197
4.799999999999998  -6.194333336233837
4.899999999999998  -6.102824057769483
4.999999999999997  -6.011270154440216
5.099999999999997  -5.9196714952765115
5.199999999999997  -5.828027993163677
5.299999999999996  -5.736339601009678
5.399999999999996  -5.644606308239029
5.499999999999996  -5.552828137582821
5.599999999999995  -5.461005142137993
5.699999999999995  -5.369137402671623
5.7999999999999945  -5.277225025148399
5.899999999999994  -5.185268138461592
5.999999999999994  -5.093266892349669
6.099999999999993  -5.001221455482535
6.199999999999993  -4.909132013702753
6.299999999999993  -4.816998768408545
6.399999999999992  -4.724821935066616
6.499999999999992  -4.632601741843886
6.599999999999992  -4.54033842834824
6.699999999999991  -4.448032244469302
6.799999999999991  -4.35568344931101
6.899999999999991  -4.263292310208558
6.99999999999999  -4.170859101822839
7.09999999999999  -4.078384105306211
7.1999999999999895  -3.9858676075338666
7.299999999999989  -3.8933099003956153
7.399999999999989  -3.8007112801433207
7.4999999999999885  -3.708072046789643
7.599999999999988  -3.6153925035540793
7.699999999999988  -3.522672956352632
7.799999999999987  -3.4299137133277737
7.899999999999987  -3.3371150844155926
7.999999999999987  -3.244277380947298
8.099999999999987  -3.1514009152824674
8.199999999999987  -3.0584860004716443
8.299999999999986  -2.9655329499460734
8.399999999999986  -2.872542077232554
8.499999999999986  -2.779513695691488
/

\setdashpattern <4pt,2pt,1pt,2pt>
\plot
1.6000000000000005  -16.41473973409694
1.7000000000000006  -14.408224520772864
1.8000000000000007  -12.85987053381083
1.9000000000000008  -11.673943663561959
2.000000000000001  -10.774046175817277
2.100000000000001  -10.09908787268282
2.200000000000001  -9.60011038336381
2.300000000000001  -9.237793300752692
2.4000000000000012  -8.980476962101353
2.5000000000000013  -8.802585213104011
2.6000000000000014  -8.683424794483214
2.7000000000000015  -8.6063708447144
2.8000000000000016  -8.55836389690933
2.9000000000000017  -8.529540873563752
3.0000000000000018  -8.512828155276042
3.100000000000002  -8.503434683915826
3.200000000000002  -8.498293687790552
3.300000000000002  -8.49554084737164
3.400000000000002  -8.494092112450335
3.500000000000002  -8.49333980326803
3.6000000000000023  -8.492953045518734
3.7000000000000024  -8.492755684271422
3.8000000000000025  -8.492655511774336
3.9000000000000026  -8.492604863656892
4.000000000000003  -8.492579325137772
4.100000000000002  -8.492566472265818
4.200000000000002  -8.492560012304635
4.300000000000002  -8.492556768433158
4.400000000000001  -8.49255514059699
4.500000000000001  -8.492554324006509
4.6000000000000005  -8.492553914503095
4.7  -8.492553709185328
4.8  -8.492553606256031
4.8999999999999995  -8.492553554660315
4.999999999999999  -8.492553528798258
5.099999999999999  -8.492553515835551
5.199999999999998  -8.492553509338485
5.299999999999998  -8.492553506082132
5.399999999999998  -8.492553504450054
5.499999999999997  -8.492553503632067
5.599999999999997  -8.492553503222098
5.699999999999997  -8.492553503016625
5.799999999999996  -8.492553502913644
5.899999999999996  -8.492553502862032
5.999999999999996  -8.492553502836165
6.099999999999995  -8.492553502823199
6.199999999999995  -8.492553502816703
6.2999999999999945  -8.492553502813445
6.399999999999994  -8.492553502811813
6.499999999999994  -8.492553502810996
6.599999999999993  -8.492553502810585
6.699999999999993  -8.49255350281038
6.799999999999993  -8.492553502810276
6.899999999999992  -8.492553502810225
6.999999999999992  -8.4925535028102
7.099999999999992  -8.492553502810187
7.199999999999991  -8.492553502810178
7.299999999999991  -8.492553502810177
7.399999999999991  -8.492553502810175
7.49999999999999  -8.492553502810175
7.59999999999999  -8.492553502810173
7.6999999999999895  -8.492553502810173
7.799999999999989  -8.492553502810173
7.899999999999989  -8.492553502810173
7.9999999999999885  -8.492553502810173
8.099999999999989  -8.492553502810173
8.199999999999989  -8.492553502810173
8.299999999999988  -8.492553502810173
8.399999999999988  -8.492553502810173
8.499999999999988  -8.492553502810173
8.599999999999987  -8.492553502810173
8.699999999999987  -8.492553502810173
8.799999999999986  -8.492553502810173
8.899999999999986  -8.492553502810173
8.999999999999986  -8.492553502810173
9.099999999999985  -8.492553502810173
9.199999999999985  -8.492553502810173
9.299999999999985  -8.492553502810173
9.399999999999984  -8.492553502810173
9.499999999999984  -8.492553502810173
9.599999999999984  -8.492553502810173
9.699999999999983  -8.492553502810173
9.799999999999983  -8.492553502810173
9.899999999999983  -8.492553502810173
9.999999999999982  -8.492553502810173
/

\setdots <3pt>
\plot
0.4  -26.648684482637986
0.5  -23.33458927564866
0.6  -20.749762339472287
0.7  -18.743247126148205
0.8  -17.19489313918617
0.9  -16.008966268937304
1.0  -15.109068781192619
1.1  -14.434110478058159
1.2  -13.935132988739149
1.3  -13.572815906128026
1.4  -13.315499567476689
1.5  -13.137607818479344
1.6  -13.018447399858548
1.7  -12.941393450089734
1.8  -12.893386502284663
1.9  -12.864563478939084
2.0  -12.847850760651374
2.1  -12.838457289291158
2.2  -12.833316293165884
2.3  -12.83056345274697
2.4  -12.829114717825668
2.5  -12.828362408643361
2.6  -12.827975650894064
2.7  -12.827778289646755
2.8  -12.827678117149668
2.9  -12.827627469032224
3.0  -12.827601930513103
3.1  -12.82758907764115
3.2  -12.827582617679967
3.3  -12.827579373808492
3.4  -12.82757774597232
3.5  -12.827576929381841
3.6  -12.827576519878427
3.7  -12.82757631456066
3.8  -12.827576211631364
3.9  -12.827576160035647
4.0  -12.82757613417359
4.1  -12.827576121210882
4.2  -12.827576114713816
4.3  -12.827576111457462
4.4  -12.827576109825387
4.5  -12.8275761090074
4.6  -12.82757610859743
4.699999999999999  -12.827576108391957
4.799999999999999  -12.827576108288977
4.899999999999999  -12.827576108237364
4.999999999999998  -12.827576108211497
5.099999999999998  -12.827576108198532
5.1999999999999975  -12.827576108192034
5.299999999999997  -12.827576108188778
5.399999999999997  -12.827576108187145
5.4999999999999964  -12.827576108186326
5.599999999999996  -12.827576108185916
5.699999999999996  -12.827576108185712
5.799999999999995  -12.827576108185609
5.899999999999995  -12.827576108185557
5.999999999999995  -12.82757610818553
6.099999999999994  -12.827576108185518
6.199999999999994  -12.827576108185513
6.299999999999994  -12.827576108185507
6.399999999999993  -12.827576108185506
6.499999999999993  -12.827576108185506
6.5999999999999925  -12.827576108185506
6.699999999999992  -12.827576108185506
6.799999999999992  -12.827576108185504
6.8999999999999915  -12.827576108185504
6.999999999999991  -12.827576108185504
7.099999999999991  -12.827576108185504
7.19999999999999  -12.827576108185504
7.29999999999999  -12.827576108185504
7.39999999999999  -12.827576108185504
7.499999999999989  -12.827576108185504
7.599999999999989  -12.827576108185504
7.699999999999989  -12.827576108185504
7.799999999999988  -12.827576108185504
7.899999999999988  -12.827576108185504
7.999999999999988  -12.827576108185504
8.099999999999987  -12.827576108185504
8.199999999999987  -12.827576108185504
8.299999999999986  -12.827576108185504
8.399999999999986  -12.827576108185504
8.499999999999986  -12.827576108185504
8.599999999999985  -12.827576108185504
8.699999999999985  -12.827576108185504
8.799999999999985  -12.827576108185504
8.899999999999984  -12.827576108185504
8.999999999999984  -12.827576108185504
9.099999999999984  -12.827576108185504
9.199999999999983  -12.827576108185504
9.299999999999983  -12.827576108185504
9.399999999999983  -12.827576108185504
9.499999999999982  -12.827576108185504
9.599999999999982  -12.827576108185504
9.699999999999982  -12.827576108185504
9.799999999999981  -12.827576108185504
9.89999999999998  -12.827576108185504
9.99999999999998  -12.827576108185504
/

\put {{$f_a=10^{11}$GeV, $m_{\tilde g}=m_{\tilde q}=1$TeV}}
[lt] <7mm,-7mm> at 0 0

\setlinear
\setdashes
\plot 7.0 -2.73976    10 -2.73976 /
\put {{$Y^{\rm EQ}_{\tilde a}\approx2\times10^{-3}$}}
[cb] <-2mm,1mm> at 8.5 -2.73976

\put {{$\tilde g$ DECAY}} [cb] <0mm,2mm> at 6 -8.492553502810173
\put {{$\chi^0_1$ DECAY}} [cb] <0mm,2mm> at 6 -12.827576108185504
\put {{SCATTERING}} [l] <10mm,0mm> at 2.2 -10.449278078633878
\setsolid
\arrow <5pt> [.2,.67] from 3.0 -10.449278078633878 to 2.2 -10.449278078633878

\endpicture$$


%% file: ckkr-fig3.tex
%
%

$$\beginpicture

\setcoordinatesystem units <30pt,27pt> point at 0 0
\setplotarea  x from -7.0 to 2.0, y from 0.0 to 10.0
\inboundscheckon
\linethickness 0.5pt

\axis bottom label {{$m_{\tilde a}$(GeV)}}
	ticks in
	width <0.8pt> length <8.0pt>
	withvalues {$10^{-6}$} {$10^{-3}$} {$1$} /
	at -6 -3 0 /
	width <0.8pt> length <8.0pt>
	at -7 -5 -4 -2 -1 1 2 /
       	width <0.5pt> length <4.0pt> at
	-6.699 -6.523 -6.398 -6.301 -6.222 -6.155 -6.097 -6.046
	-5.699 -5.523 -5.398 -5.301 -5.222 -5.155 -5.097 -5.046
	-4.699 -4.523 -4.398 -4.301 -4.222 -4.155 -4.097 -4.046
	-3.699 -3.523 -3.398 -3.301 -3.222 -3.155 -3.097 -3.046
	-2.699 -2.523 -2.398 -2.301 -2.222 -2.155 -2.097 -2.046
	-1.699 -1.523 -1.398 -1.301 -1.222 -1.155 -1.097 -1.046
	-0.699 -0.523 -0.398 -0.301 -0.222 -0.155 -0.097 -0.046
       	0.301 0.477 0.602 0.699 0.778 0.845 0.903 0.954
       	1.301 1.477 1.602 1.699 1.778 1.845 1.903 1.954
       	/
/

\axis top label {}
	ticks in
	width <0.8pt> length <8.0pt>
	at -7 -6 -5 -4 -3 -2 -1 0 1 2 /
       	width <0.5pt> length <4.0pt> at
	-6.699 -6.523 -6.398 -6.301 -6.222 -6.155 -6.097 -6.046
	-5.699 -5.523 -5.398 -5.301 -5.222 -5.155 -5.097 -5.046
	-4.699 -4.523 -4.398 -4.301 -4.222 -4.155 -4.097 -4.046
	-3.699 -3.523 -3.398 -3.301 -3.222 -3.155 -3.097 -3.046
	-2.699 -2.523 -2.398 -2.301 -2.222 -2.155 -2.097 -2.046
	-1.699 -1.523 -1.398 -1.301 -1.222 -1.155 -1.097 -1.046
	-0.699 -0.523 -0.398 -0.301 -0.222 -0.155 -0.097 -0.046
       	0.301 0.477 0.602 0.699 0.778 0.845 0.903 0.954
       	1.301 1.477 1.602 1.699 1.778 1.845 1.903 1.954
       	/
/

\axis left label {{$T_R$(GeV)}}
        ticks in
        width <0.8pt> length <8.0pt>
        withvalues {$1$} {$10^3$} {$10^6$} {$10^9$} /
        at 0 3 6 9 /
        width <0.8pt> length <8.0pt>
        at 1 2 4 5 7 8 /
        width <0.5pt> length <4.0pt> at
        0.301 0.477 0.602 0.699 0.778 0.845 0.903 0.954
        1.301 1.477 1.602 1.699 1.778 1.845 1.903 1.954
        2.301 2.477 2.602 2.699 2.778 2.845 2.903 2.954
        3.301 3.477 3.602 3.699 3.778 3.845 3.903 3.954
        4.301 4.477 4.602 4.699 4.778 4.845 4.903 4.954
        5.301 5.477 5.602 5.699 5.778 5.845 5.903 5.954
        6.301 6.477 6.602 6.699 6.778 6.845 6.903 6.954
        7.301 7.477 7.602 7.699 7.778 7.845 7.903 7.954
        8.301 8.477 8.602 8.699 8.778 8.845 8.903 8.954
        9.301 9.477 9.602 9.699 9.778 9.845 9.903 9.954
        /
/

\axis right label {}
        ticks in
        width <0.8pt> length <8.0pt>
        at 1 2 3 4 5 6 7 8 9 /
        width <0.5pt> length <4.0pt> at
        0.301 0.477 0.602 0.699 0.778 0.845 0.903 0.954
        1.301 1.477 1.602 1.699 1.778 1.845 1.903 1.954
        2.301 2.477 2.602 2.699 2.778 2.845 2.903 2.954
        3.301 3.477 3.602 3.699 3.778 3.845 3.903 3.954
        4.301 4.477 4.602 4.699 4.778 4.845 4.903 4.954
        5.301 5.477 5.602 5.699 5.778 5.845 5.903 5.954
        6.301 6.477 6.602 6.699 6.778 6.845 6.903 6.954
        7.301 7.477 7.602 7.699 7.778 7.845 7.903 7.954
        8.301 8.477 8.602 8.699 8.778 8.845 8.903 8.954
        9.301 9.477 9.602 9.699 9.778 9.845 9.903 9.954
        /
/

\setplotsymbol (.)

\setquadratic

\setsolid
\plot
10.5497966703	1.5
7.9649697341	1.6
5.95845452077	1.7
4.41010053381	1.8
3.20263090111	1.9
2.29480532719	2.0
1.60842685711	2.1
1.09285506263	2.2
0.70644652515	2.3
0.414179184433	2.4
0.18689218128	2.5
0.001571734156	2.6
-0.17		2.7
-0.34		2.8
-0.48		2.9
-0.61533971889	3.0
-0.70615428438	3.1
-0.79699601925	3.2
-0.88786718682	3.3
-0.97876981350	3.4
-1.06970571112	3.5
-1.16067649699	3.6
-1.25168361194	3.7
-1.34272833648	3.8
-1.43381180546	3.9
-1.52493502123	4.0
-1.61609886555	4.1
-1.70730411039	4.2
-1.79855142766	4.3
-1.88984139808	4.4
-1.98117451921	4.5
-2.07255121273	4.6
-2.16397183115	4.7
-2.25543666377	4.8
-2.34694594223	4.9
-2.43849984556	5.0
-2.53009850472	5.1
-2.62174200684	5.2
-2.71343039899	5.3
-2.80516369176	5.4
-2.89694186242	5.5
-2.98876485786	5.6
-3.08063259733	5.7
-3.17254497485	5.8
-3.26450186154	5.9
-3.35650310765	6.0
-3.44854854452	6.1
-3.5406379863	6.2
-3.63277123159	6.3
-3.72494806493	6.4
-3.81716825816	6.5
-3.90943157165	6.6
-4.00173775553	6.7
-4.09408655069	6.8
-4.18647768979	6.9
-4.27891089818	7.0
-4.37138589469	7.1
-4.46390239247	7.2
-4.5564600996 	7.3
-4.64905871986	7.4
-4.74169795321	7.5
-4.83437749645	7.6
-4.92709704365	7.7
-5.01985628667	7.8
-5.11265491558	7.9
-5.20549261905	8.0
-5.29836908472	8.1
-5.39128399953	8.2
-5.48423705005	8.3
-5.57722792277	8.4
-5.67025630431	8.5
/

\setlinear
\setdashes
\plot -5.7101 8.5   -5.7101 10 /


\setshadesymbol <z,z,z,z> ({.})
\setshadegrid span <0.8pt>
\setquadratic
\hshade
1.0	-1.7	2
1.1	-1.7	2
1.2	-1.7	2
1.3	-1.7	2
1.4	-1.7	2
1.5	-1.7	2
1.6	-1.7	2
1.7	-1.7	2
1.8	-1.7	2
1.9	-1.7	2
2.0	-1.7	2
2.1	-1.7	1.60842685711  
2.2	-1.7	1.09285506263  
2.3	-1.7	0.70644652515  
2.4	-1.7	0.414179184433  
2.5	-1.7	0.18689218128  
2.6	-1.7	0.00157173415649  
3.0	-1.7	-0.615339718895  
3.1	-1.7	-0.706154284386  
3.2	-1.7	-0.796996019255  
3.3	-1.7	-0.887867186825  
3.4	-1.7	-0.978769813502  
3.5	-1.7	-1.06970571112  
3.6	-1.7	-1.16067649699  
3.7	-1.7	-1.25168361194  
3.8	-1.7	-1.34272833648  
3.9	-1.7	-1.43381180546  
4.0	-1.7	-1.52493502123  
4.1	-1.7	-1.61609886555  
4.2	-1.7	-1.70730411039  
4.3	-1.7	-1.79855142766  
4.4	-1.7	-1.88984139808  
4.5	-1.7	-1.98117451921  
/
\setshadesymbol <z,z,z,z> ({\tiny.})
\setshadegrid span <1.5pt>
\setlinear
\hshade
0.7	-2.0	2
1.0	-2.0	2
/
\hshade
1.0	-2.0	-1.7
4.0	-2.0	-1.7
/
\setquadratic
\hshade
4.0	-2.0	-1.7
4.1	-2.0	-1.61609886555
4.2	-2.0	-1.70730411039
4.3	-2.0	-1.79855142766
4.4	-2.0	-1.88984139808
4.5	-2.0	-1.98117451921
4.51	-2.0	-1.98117451921
/

\begingroup\color{white}
\put {{$\Omega^{\rm NTP}_{\tilde a}h^2\approx1$}}
[c] <0mm,0mm> at 0 1.7
\endgroup

\setsolid
\setlinear
\plot -2 0  -2 5 /
\multiput {$\backslash$} [rb] <1pt,0pt> at -2 0 *23 0.0 0.2 /

\setdashes
\setlinear
\plot -3 0.7  2.0 0.7 /
\put {{$\displaystyle T_F\approx\frac{m_\chi}{20}$}}
[l] <2mm,0mm> at 2.0 0.7

\put {{$\Omega^{\rm TP}_{\tilde a}h^2=1$}} [c] <0mm,-3mm> at -4 5.5
\setsolid
\arrow <5pt> [.2,.67] from -4.0 5.5 to -3.5406379863 6.2

\put {{EXCLUDED}} [c] <0mm,+3mm> at -1 7
\put {{($\Omega^{\rm TP}_{\tilde a}h^2>1$)}} [c] <0mm,-3mm> at -1 7

\setlinear
\setdashes
\plot -5.7 0  -5.7 2.5 /
\plot -4.0 0  -4.0 2.5 /
\put {{HOT}} [c] <0mm,0mm> at -6.3 1.7
\put {{WARM}} [c] <0mm,0mm> at -4.85 1.7
\put {{COLD}} [l] <2mm,0mm> at -4.0 1.7
\setsolid
\arrow <5pt> [.2,.67] from -4.0 2.1 to -3.0 2.1

\endpicture$$


%% file: cosmo2k+idm2k-lr-net.bbl
\begin{thebibliography}{99}
\def\apj#1#2#3{Astrophys.\ J.\ {\bf #1}, #2 (#3)}
\def\ijmp#1#2#3{Int.\ J.\ Mod.\ Phys.\ {\bf #1}, #2 (#3)}
\def\mpl#1#2#3{Mod.\ Phys.\ Lett.\ {\bf #1}, #2 (#3)}
\def\nat#1#2#3{Nature\ {\bf #1}, #2 (#3)}
\def\npb#1#2#3{Nucl.\ Phys.\ {\bf B#1}, #2 (#3)}

\def\plb#1#2#3{Phys.\ Lett.\ {\bf B#1}, #2 (#3)}
\def\prd#1#2#3{Phys.\ Rev.\ {\bf D#1}, #2 (#3)}
\def\prl#1#2#3{Phys.\ Rev.\ Lett.\ {\bf #1}, #2 (#3)}
\def\prt#1#2#3{Phys.\ Rep.\ {\bf #1}, #2 (#3)}
\def\sjnp#1#2#3{Sov.\ J.\ Nucl.\ Phys.\ {\bf #1}, #2 (#3)}
\def\zp#1#2#3{Z.\ Phys.\ {\bf #1}, #2 (#3)}

\bibitem{ckr} L. Covi, J.E. Kim and L. Roszkowski,
  \prl{82}{4180}{1999}.

\bibitem{ckkr} L. Covi, H.B. Kim, 
J.E. Kim and L. Roszkowski, 
hep-ph/0101009.

\bibitem{ksvz}
J.E. Kim, \prl{43}{103}{1979};
M.A. Shifman, V.I. Vainstein, and V.I. Zakharov, \npb{166}{4933}{1980}.

\bibitem{dfsz} 
M. Dine, W. Fischler, and M. Srednicki, \plb{104}{99}{1981};
A.P. Zhitnitskii, \sjnp{31}{260}{1980}.

\bibitem{axionreviews:cite}
J.E. Kim, \prt{150}{1}{1987};
M.S. Turner, \prt{197}{67}{1990};
G.G. Raffelt,  \prt{198}{1}{1990};
P. Sikivie, hep-ph/0002154.
  
\bibitem{tw}
K. Tamvakis and D. Wyler, \plb{112}{451}{1982}.

\bibitem{rtw} 
K. Rajagopal, M. S. Turner, and F. Wilczek, \npb{358}{447}{1991}.

\bibitem{ckn}
E.J. Chun, J.E. Kim and H. P. Nilles, \plb{287}{123}{1992}.

\bibitem{na93} P.~Nath and 
R.~Arnowitt, \prl{69}{725}{92}.
\bibitem{rr93} 
R.G. Roberts and L. Roszkowski, \plb{309}{329}{1993}.
\bibitem{kkrw}
G.L. Kane, C. Kolda, L. Roszkowski, and J.D. Wells, \prd{49}{6173}{1994}.

\bibitem{bbnbound} 
J. Ellis, \etal, \npb{373}{399}{1992}. 

\bibitem{gkr00}
G.F. Giudice, E.W. Kolb, and A. Riotto, hep-ph/0005123.

\end{thebibliography}
